\documentclass[12pt]{article}
\usepackage{amsmath}
\usepackage{amssymb}
\usepackage{graphicx}
\usepackage{float}
\usepackage{indentfirst}
\usepackage{mathrsfs}
\usepackage{dsfont}

\usepackage{setspace}

\usepackage[labelfont=bf,labelsep=period]{caption}

\usepackage[numbers,sort&compress]{natbib}

\newtheorem{theorem}{Theorem}
\newtheorem{proposition}{Proposition}
\newtheorem{corollary}{Corollary}
\newtheorem{observation}{Observation}

\title{The Generalized Uncertainty Principle}

\author
{Jun-Li Li and Cong-Feng Qiao$^{\ast}$ \\ [0.2cm]
\normalsize{School of Physical Sciences, University of Chinese Academy of Sciences} \\
\normalsize{YuQuan Road 19A, Beijing 100049, China}\\
\normalsize{Center of Materials Science and Optoelectronics Engineering \& CMSOT,} \\
\normalsize{University of Chinese Academy of Sciences, YuQuan Road 19A, Beijing 100049, China}\\
\normalsize{Key Laboratory of Vacuum Physics, University of Chinese Academy of Sciences} \\
\normalsize{YuQuan Road 19A, Beijing 100049, China}\\
\normalsize{$^\ast$ To whom correspondence should be addressed; E-mail: qiaocf@ucas.ac.cn.}
}

\date{}

\begin{document}
\baselineskip24pt \maketitle

\begin{abstract} \doublespacing
The uncertainty principle lies at the heart of quantum physics, and is widely thought of as a fundamental limit on the measurement precisions of incompatible observables. Here we show that the traditional uncertainty relation in fact belongs to the leading order approximation of a generalized uncertainty relation. That is, the leading order linear dependence of observables gives the Heisenberg type of uncertainty relations, while higher order nonlinear dependence may reveal more different and interesting correlation properties. Applications of the generalized uncertainty relation and the high order nonlinear dependence between observables in quantum information science are also discussed.
\end{abstract}

\newpage

\section{Introduction}

The proposition of the uncertainty principle may be attributed to the early efforts devoted to incorporate the wave and particle natures to each individual quantum. Heisenberg stated that the canonically conjugate quantities, $x$ and $p$, can be determined simultaneously only with a characteristic indeterminacy \cite{Heisenberg-1}. A well known formulation of the uncertainty relation takes the following form \cite{Robertson-1}
\begin{align}
\Delta X^2 \Delta Y^2  \geq \frac{1}{4} \left|\langle [X,Y] \rangle \right|^2 \; , \label{Uncertainty-Hei}
\end{align}
where the variance $\Delta X^2 \equiv \langle X^2 \rangle - \langle X \rangle^2 $ is a measure of the uncertainty for observable $X$ (similarly for $Y$) and $[X,Y] \equiv XY-YX$ is the commutator. In relation (\ref{Uncertainty-Hei}), $X$ and $Y$ are no longer restricted to canonical variables but can be arbitrary observables. An improvement of the  uncertainty relation was made by Schr\"odinger \cite{Schrodinger-1}
\begin{align}
\Delta X^2 \Delta Y^2  \geq \frac{1}{4} \left| \langle [X,Y]\rangle \right|^2 + \frac{1}{4} \left| \langle \{X,Y\} \rangle - 2\langle X\rangle \langle Y \rangle \right|^2 \; . \label{Uncertainty-Sch}
\end{align}
Here $\{X,Y\} \equiv XY+YX$ is the anticommutator. The uncertainty relations (\ref{Uncertainty-Hei}) and (\ref{Uncertainty-Sch}) are normally explained as trade-off relations for the uncertainties of incompatible observables, which is lower bounded by the expectation values of their (anti)commutator. In other words, the variances of incompatible observables are not independent but correlated with each other. In recent years, important progresses have been made in the study of variance based uncertainty relations, e.g. \cite{Pati-1, Bloch-Va}, but few studies concern about the higher order moments of observables.

The lower bounds of the above variance based uncertainty relations depend on the quantum state and may reach zero, in which case the uncertainty relations become trivial. Partly due to this reason, the entropic uncertainty relation was introduced \cite{Entropy-0}, with a typical form of \cite{Entropy-1}
\begin{equation}
H(X)+H(Y) \geq \ln \left(\frac{1}{c} \right)\; ,
\end{equation}
where $c=\max_{i,j} |\langle x_i|y_j\rangle|^2$ is the maximum overlap of eigenbases $|x_i\rangle$ and $|y_j\rangle$ of $X$ and $Y$ respectively and it turns out to be state independent. $\ln(\cdot)$ here is the natural logarithm. The Shannon entropy $H(\cdot)$ is another measure of uncertainty for observables regarding to the probability distribution of the measurement outcomes. In the presence of quantum memory, the bound of the entropic unceratinty relation may be reduced \cite{Entropy-2}, see Refs. \cite{Entropy-3} and \cite{Entropy-4} for recent developments. Note, recent study indicates that these two superficially different forms of uncertainty relations (variance based and entropic) are in fact interconvertible \cite{Equiv-va-en}. Moreover, it has been realized that the variance or entropy alone may not be sufficient to fully characterize the properties of quantum uncertainty \cite{Maj-Latt-Un}, higher order moments of observables tend to be nontrivial \cite{Equiv-va-en}. The variance is merely the second-order central moment.

In this work, by interpreting the incompatibility as the statistical dependence between observables, we derive a generalized uncertainty relation (GUR) capable of embodying the incompatibility (dependence) of observables to arbitrary order. In Section 2, after a reinterpretation of complementary we employ the statistical quantity of cumulant to measure the dependence between observables in different orders. Explicit form of GUR is derived and expanded in terms of cumulants in Section 3. The first few terms in the expansion are studied in the subsections of Section 3. The leading order nontrivial linear dependence turns out to be the Heisenberg type uncertainty relation. The first nonlinear dependence between incompatible observables gives the ``Skewness uncertainty relation''. In Section 4, concrete examples and simple applications of the new findings are given for understanding the high order dependence. Section 5 remains for conclusion.

\section{The uncertainty principle and dependence}

\begin{figure}\centering
\scalebox{0.5}{\includegraphics{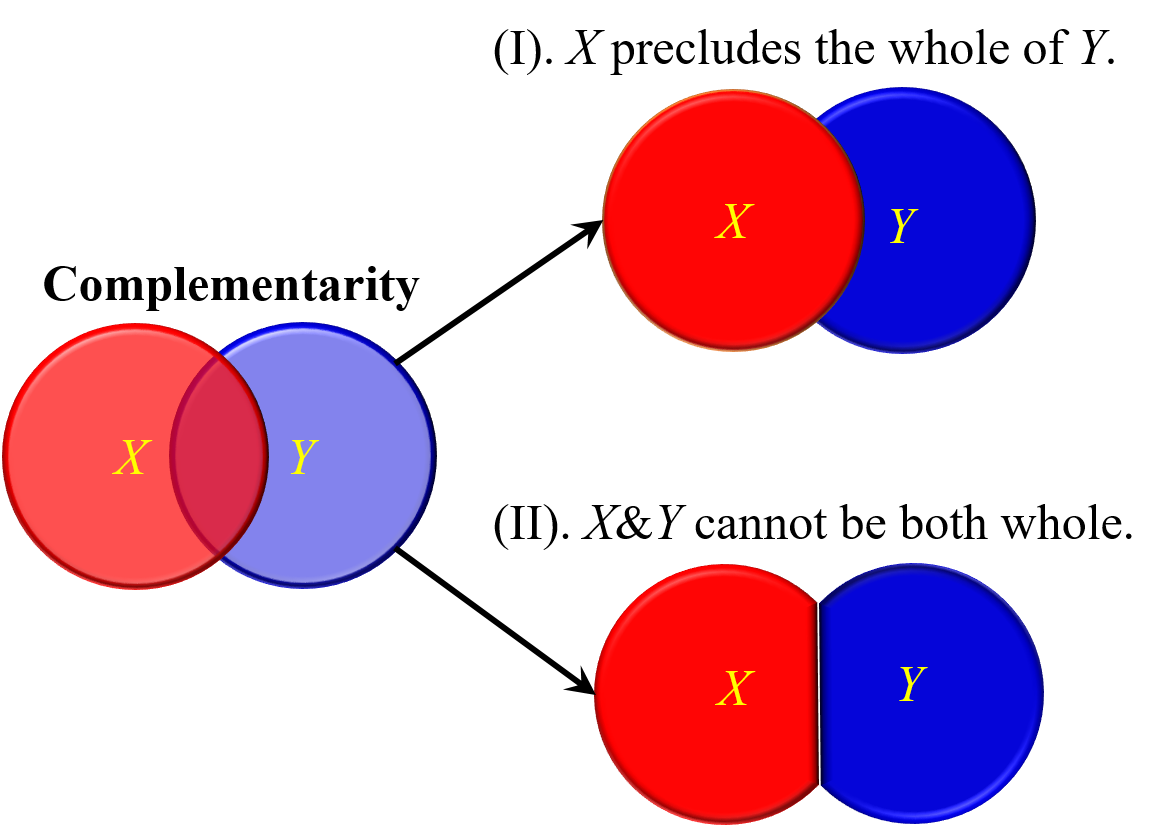}}
\caption{{\bf Two interpretations of complementarity.} Regarding $X\cup Y$ as a whole, we have that: (I) the completeness of $X$ precludes the wholeness of $Y$; (II) $X$ and $Y$ cannot be as a whole simultaneously.} \label{Fig-Complementary}
\end{figure}

To understand the wave-particle duality in atomic phenomena, a fundamental concept ``Complementarity Principle'' was proposed by Bohr, which may be stated as ``any given application of classical concepts precludes the simultaneous use of other classical concepts which in a different connection are equally necessary for the elucidation of the phenomena'' \cite{Complementary-B}. While Heisenberg put forward a more operational idea ``that canonically conjugate quantities can be determined simultaneously only with a characteristic indeterminacy'' \cite{Heisenberg-1}. These two statements about the complementarity of two incompatible observables, $X$ and $Y$, are illustrated in Figure \ref{Fig-Complementary}. While Bohr's interpretation implies that the precise determination of one observable precludes the other, the Heisenberg's interpretation reflects the fact that the two observables cannot be determined simultaneously. Note that both Heisenberg's uncertainty principle and Bohr's interpretation concern about the impaired sections in measurement. In view of the overlapped section, we are encouraged to think of a slightly different interpretation for the complementarity principle,
\begin{observation}
{\bf The generalized uncertainty principle:} In quantum theory, if the quantities are incompatible, they are mutually dependent. \label{observation-1}
\end{observation}
The generalized uncertainty principle indicates that measuring one observable may provide you some information about its incompatible partners. To ascertain its physical consequences, we need to quantify the generalized uncertainty principle. For this purpose, we first define the statistical dependence of physical observables.

Given a random variable $X$, the moment generating function takes the following form
\begin{align}
\langle e^{sX}\rangle =\sum_{n=0}^{\infty}\langle X^n\rangle \frac{s^n}{n!} \; ,\; s\in \mathbb{C}\; . \label{Moment-Gen}
\end{align}
Here $\langle X\rangle$ means the expectation value of a variable $X$ and the parameter $s$ is a complex number. The logarithm (with base $e$ if not specified) of equation (\ref{Moment-Gen}) then generates the cumulants \cite{Adv-statistics}, that is
\begin{align}
K(sX) & \equiv \log(\langle e^{sX}\rangle)= \log\left(1+s\langle X\rangle + \frac{s^2}{2!}\langle X^2\rangle + \frac{s^3}{3!}\langle X^3\rangle + \cdots \right) \nonumber \\
& = \sum_{m=1}^{\infty} \frac{s^m}{m!}\kappa_m(X) \; , \label{Kx-Expansion}
\end{align}
where the sum runs over a power series of $s$ whose coefficients $\kappa_m(X)$ are collections of different orders of moments. $\kappa_m(X)$ is called the cumulant of order $m$ which exists if the $m$th and lower orders of moments of $X$ exist \cite{Adv-statistics}. The first few orders of cumulants are
\begin{eqnarray}
\mathrm{Mean} & : & \kappa_1  = \langle X \rangle \; ,   \label{Cumulant-X-1}\\
\mathrm{Variance} & : &  \kappa_2   = \langle X^2 \rangle - \langle X\rangle^2 \; , \label{Cumulant-X-2} \\
\mathrm{Skewness} & : & \kappa_3  = \langle X^3 \rangle - 3 \langle X^2 \rangle \langle X \rangle + 2 \langle X \rangle^3\; ,  \label{Cumulant-X-3}\\
\mathrm{Kurtosis} & : & \kappa_4  = \langle X^4 \rangle - 4 \langle X^3 \rangle \langle X \rangle - 3 \langle X^2 \rangle^2 +12 \langle X^2\rangle \langle X \rangle^2 -6 \langle X \rangle^4\; . \label{Cumulant-X-4}
\end{eqnarray}
Here $\kappa_1$ is the expectation value of an observable with given distribution; the variance $\kappa_2$ measures the spread of the distribution; the skewness $\kappa_3$ reflects the distribution asymmetry; the kurtosis $\kappa_4$ measures the tailedness.

For two random variables $X$ and $Y$, the cumulants generating function is
\begin{align}
K(sX+tY) \equiv \log(\langle e^{sX+tY} \rangle) = \sum_{m+n=1}^{\infty} \kappa_{mn} \frac{s^mt^n}{m!n!} \; , \label{Kxy-expansion}
\end{align}
where $\kappa_{mn}$ are named cross cumulants \cite{Three-lect}. While $\kappa_{m0}$ and $\kappa_{0n}$ have similar expressions as equations (\ref{Cumulant-X-1})-(\ref{Cumulant-X-4}), the first two terms of cross cumulants $\kappa_{mn}$ read
\begin{align}
\kappa_{11} & =  \frac{1}{2}\langle\{X,Y\}\rangle - \langle X \rangle\langle Y\rangle \; , \label{Cross-cumulant-11}\\
\kappa_{12} & =  \frac{1}{3}\langle \{X,Y,Y\}\rangle - (\langle X\rangle\langle Y^2\rangle + \langle\{X,Y\}\rangle \langle Y\rangle) + 2\langle X\rangle \langle Y\rangle^2 \; . \label{Cross-cumulant-12}
\end{align}
Here $\{X,Y,Y\}\equiv XYY+YXY+YYX$ is defined as the 3rd order anticommutator and subscripts in $\kappa_{mn}$ indicate that there are $m$ $X$s and $n$ Ys in the expansion of moments. The cross cumulants for multiple variables can be similarly defined
\begin{align}
K(s_1X_1+s_2X_2+\cdots+ s_NX_N) \equiv \log( \langle e^{\vec{s}\cdot\vec{X}}\rangle ) \; .
\end{align}
According to the Corollary of Theorem I in Ref. \cite{Kubo-1962} we further have:
\begin{observation}
The cross cumulant is nonzero, if and only if its variables are statistically connected. \label{Observation-2}
\end{observation}
Observation \ref{Observation-2} implies that, the cumulants are capable of quantifying the statistical dependence of physical observables. Considering the simplest non-trivial case of two variables $X$ and $Y$, there exists \cite{Three-lect}
\begin{corollary}
The cumulants linearize addition of independent random variables
\begin{align}
X,Y \; \text{independent} \Longrightarrow \kappa_n(sX + tY) -[ s^n\kappa_n(X) + t^n\kappa_n(Y)] =0 \; , \forall n \geq 1 \; ,
\end{align}
where we have used the fact $\kappa_n(sX)=s^n \kappa_n(X)$. \label{Corollary-linearization}
\end{corollary}
The contrapositive of Corollary \ref{Corollary-linearization} is: if there is any $n$ that $\kappa_n(sX+tY) -[\kappa_n(sX)+ \kappa_n(tY)] \neq 0$, then $X$ and $Y$ are $n$th order dependent. That is, the difference between the cumulant of the sums and the sum of the cumulants signifies the statistical dependence between observables.

\section{The generalized uncertainty relation}

Observation \ref{Observation-2} indicates that the dependence in the generalized uncertainty principle may be demonstrated by the cross cumulants which are generated via the expansion of exponential operator sums. Consider the Cauchy-Schwarz inequality for the two state vectors of $e^{s^*X}|\psi\rangle$ and $e^{tY}|\psi\rangle$, we have the following GUR:
\begin{theorem}
For arbitrary observables $X$ and $Y$, there exists a generalized uncertainty relation
\begin{align}
K[(s+s^*)X] + K[(t+t^*)Y] \geq K(Z_{st}) +  K^*(Z_{st})  \; , \; s, t \in \mathbb{C} \; . \label{K-uncertainty}
\end{align}
Here $K(\cdot)$ signifies the generating function of cumulants defined in equation (\ref{Kx-Expansion}); * means the complex conjugation; $Z_{st}= \log(e^{sX}e^{tY}) = Z_{1}+ Z_{11}+\cdots$ is defined as
\begin{align}
Z_1 = sX +tY\; , \; Z_{11} = \frac{1}{2}[sX,tY]\;, \cdots \; ,\label{Haus-Z}
\end{align}
in light of the well-known Baker-Campbell-Hausdorff (BCH) formula.
\end{theorem}
The proof of Theorem 1 is presented in the Appendix A. The primary merit of equation (\ref{K-uncertainty}) is that the quantum and classical (say communitive hereafter) dependence between observables can be distinguished via high order powers of $s,t$ in $Z_{st}$, i.e., the commutators in the BCH formula. For commutative observables, equation (\ref{K-uncertainty}) predicts
\begin{align}
K[(s+s^*)X] + K[(t+t^*)Y] \geq K(sX+tY)+ \mathrm{c.c. }\; , \label{Comm-indep-XY}
\end{align}
where ``c.c.'' means the complex conjugation of the previous term. The GUR of equation (\ref{K-uncertainty}) then may be expressed in a more distinct form
\begin{align}
& \left\{ K[(s+s^*)X] -K(sX) -K^*(sX) \right\} + \left\{ K[(t+t^*)Y] -K(tY)-K^*(tY) \right\} \nonumber \\ \geq &  \left[ K(Z_{st}) -K(sX)-K(tY) \right] + \mathrm{c.c.}  \; . \label{GUR-familiar}
\end{align}
The right hand side of inequality (\ref{GUR-familiar}) is composed of (anti)commutators of different orders and reveals the dependence between observables according to Corollary \ref{Corollary-linearization}. We may say that, the sum of the statistical properties of each observable on the left hand side of inequality (\ref{GUR-familiar}) is lower bounded by their statistical dependence on the right hand side.  The dependence would be zero for observables without correlations. We shall show what new the GUR may tell by expanding it in $s$ and $t$ in the infinitesimal limit.

\subsection{The first order: Trivial identity for mean values}

The first order cumulant is the mean value. Comparing the coefficients of $s$, $t$, and their complex conjugates on both sides of equation (\ref{K-uncertainty}) gives
\begin{eqnarray}
(s+s^*)\langle X\rangle + (t +t^*) \langle Y\rangle =  \langle sX+tY \rangle + \langle sX+tY \rangle^* \; . \label{first-order}
\end{eqnarray}
The equality (\ref{first-order}) is established on: 1. The expectation values of Hermitian operators are real, $\langle X\rangle=\langle X\rangle^*$; 2. The superposition principle holds for Hermitian operators, $\langle X+Y\rangle = \langle X\rangle + \langle Y\rangle$. The equality means that there is no contribution from the mean value to the GUR.

\subsection{The second order: Variance uncertainty relations}

Expanding equation (\ref{GUR-familiar}) to the second order and neglecting the high order contributions of $O(s^mt^n)$ for $n+m\geq 3$, we have
\begin{corollary}
For two observables $X$ and $Y$, there exists the following uncertainty relation for cumulant $\kappa_2$
\begin{equation}
|s|^2 \kappa_2(X) + |t|^2 \kappa_2(Y) \geq\left[ \kappa_{11}(sX,tY) + \langle Z_{11}\rangle \right] + \mathrm{c.c.} \; ,\label{W-uncertainty-2rd}
\end{equation}
where $s,t \in \mathbb{C}$ and $|s|, |t| \ll 1$. Note, $\langle Z_{11}\rangle = \displaystyle \frac{st}{2}\langle [X,Y] \rangle$ may not be purely imaginary for complex $s$ and $t$. \label{Corollary-variance}
\end{corollary}
The proof Corollary \ref{Corollary-variance} is presented in Appendix B. The right hand side of equation (\ref{W-uncertainty-2rd}) is composed of the expectation values of a commutator and an anticommutator (see equation (\ref{Cross-cumulant-11})) and reflects the linear dependence between $X$ and $Y$. To exemplify this, we write the second order cumulants in terms of variance
\begin{align}
|s|^2\Delta X^2 +|t|^2\Delta Y^2 & \geq |st| \sqrt{\left| \langle [X,Y]\rangle \right|^2 + \left|\langle \{X,Y\}\rangle -2\langle X\rangle\langle Y\rangle \right|^2} \nonumber \\
& = 2|st||\langle (X - \langle X\rangle)(Y-\langle Y\rangle)\rangle|\;. \label{variance-Pearson}
\end{align}
Let $|s| = \varepsilon \sqrt{\frac{\Delta Y}{\Delta X}}$ and $|t| = \varepsilon \sqrt{\frac{\Delta X}{\Delta Y}}$, relation (\ref{variance-Pearson}) implies the uncertainty relation (\ref{Uncertainty-Sch}), and gives out a constraint on Pearson correlation coefficient $\rho_{X,Y}$ which is a measure of correlation (linear dependence) between two variables $X$ and $Y$
\begin{align}
\rho_{X,Y} = \frac{\left|\langle (X-\langle X\rangle)(Y-\langle Y\rangle)\rangle \right|}{\Delta X\Delta Y} \leq 1 \; . \label{P-corr-coeff}
\end{align}
From equation (\ref{P-corr-coeff}) we know that the maximal dependence occurs when $(X-\langle X\rangle) |\psi\rangle \propto (Y-\langle Y\rangle)|\psi\rangle$, viz. $\rho_{X,Y}=1$.

We may regard any nontrivial statistical constraint on two or more correlated physical observables as the uncertainty relation, i.e., equation (\ref{W-uncertainty-2rd}) in Corollary \ref{Corollary-variance} is an uncertainty relation for the statistical quantity of variance (the 2nd order cumulant). The GUR exhibits the statistical constraint between observables in full order dependence. In this sense, the GUR may be regarded as a superset of inequivalent uncertainty relations, which is applicable to both classical and quantum quantities. According to (\ref{W-uncertainty-2rd}), the uncertainty relation (UR) for classical commutative observables and quantum UR for non-commutative observables then turn to be (up to the second order)
\begin{eqnarray}
\text{Classical UR} &:& |s|^2 \kappa_2(X) + |t|^2 \kappa_2(Y)  \geq  \kappa_{11}(sX,tY) + \mathrm{c.c.} \; ,\\
\text{Quantum UR} &:& |s|^2 \kappa_2(X) + |t|^2 \kappa_2(Y)  \geq\left[ \kappa_{11}(sX,tY) + \langle Z_{11}\rangle \right] + \mathrm{c.c.} \; .
\end{eqnarray}
When the cross cumulant $\kappa_{11}=0$ the classical theory predicts no constraint on variances, while in quantum mechanics there remains a constraint induced by $Z_{11}$ (here $\langle Z_{11}\rangle$ is not purely imaginary). In the case of $\kappa_{11}=0$ and $\langle Z_{11}\rangle=0$, the Heisenberg uncertainty relation would become trivial, whereas the GUR remains meaningful since hereby $X$ and $Y$ are only linearly uncorrelated but may be nonlinearly dependent, i.e., nonlinear dependence may appear in high order expansions.

\subsection{The third order: Skewness uncertainty relation}

\begin{figure}\centering
\scalebox{0.5}{\includegraphics{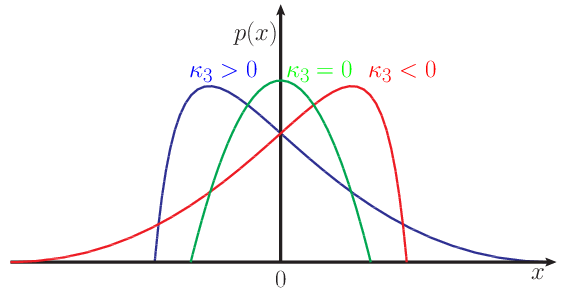}}
\caption{{\bf The skewness of different distributions.} Given a probabiity distribution function $p(x)$ of the random variable $X$, the skewness $\kappa_3$ describes the distribution asymmetry. $\kappa_3$ is zero when distribution is symmetric.} \label{Fig-Skew-Kurt}
\end{figure}

While correlation is broadly employed to indicate the statistical linear dependence of random variables, GUR enables us to explore the higher order nonlinear dependence, a peculiar character of GUR. Expand equation (\ref{K-uncertainty}) or (\ref{GUR-familiar}) to the third order, i.e., keeping those $s^mt^n$ terms with $m+n\leq 3$ while neglecting the higher order ones in the small $s$ and $t$ limits, we have the following Corollary:
\begin{corollary}
For two observables $X$ and $Y$, following skewness uncertainty relation holds:
\begin{align}
& |s|^2\left[\kappa_{2}(X) +\frac{s+s^*}{2} \kappa_3(X) \right]+
|t|^2\left[\kappa_{2}(Y) +\frac{t+t^*}{2} \kappa_3(Y) \right] \nonumber \\
& \geq \left[\kappa_{11}(sX,tY)+\frac{\kappa_{12}(sX,tY)+\kappa_{21}(sX,tY)}{2} + \right. \nonumber \\
& \hspace{0.5cm} \left. \langle Z_{11}+ Z_{12}+ Z_{21}\rangle + \frac{\langle \{ Z_{1},Z_{11}\} \rangle - 2\langle Z_1\rangle\langle Z_{11}\rangle}{2!} \right] + \mathrm{c.c.}  \; . \label{Skew-uncetainty}
\end{align}
Here $s,t \in \mathbb{C}$, $|s|, |t| \ll 1$; $Z_1=sX+tY$, $Z_{11} =  \frac{1}{2}[sX,tY]$, $Z_{21} = \frac{1}{12}[sX, [sX, tY]]$, and $Z_{12} = \frac{1}{12}  [tY, [tY, sX]]$. \label{Corollary-2}
\end{corollary}
The procedure on how to truncate power series to the desired order is demonstrated in the Appendix C. The third order cumulant $\kappa_3$ names ``skewness'', which characterizes the distribution asymmetry, as shown in Figure \ref{Fig-Skew-Kurt}.

In a special case of $s=t=\varepsilon$ being real and $[X,Y]=0$, equation (\ref{Skew-uncetainty}) gives the following variance-skewness relation
\begin{align}
& \kappa_2(X) + \kappa_2(Y) +   \varepsilon \left[ \kappa_3(X)+ \kappa_3(Y) \right] \nonumber \\
\geq & \langle \{X,Y\} \rangle - 2\langle X\rangle \langle Y\rangle +\nonumber \\
& \varepsilon \left[ \frac{1}{3} \langle\{X,Y,Y\} \rangle -(\langle X\rangle \langle Y^2\rangle + \langle \{X,Y\}\rangle\langle Y\rangle)+2\langle X\rangle \langle Y\rangle^2 + (X\leftrightarrow Y)\right] \;, \label{Three-O-C}
\end{align}
where $X\leftrightarrow Y$ means the exchange of the observables in previous terms. The right hand side of equation (\ref{Three-O-C}) is zero for independent observables $X$ and $Y$.

\section{The full order dependence and examples}

In view of the second and third order expansions of GUR, it is clear that there are two types of terms on the right hand side of (\ref{GUR-familiar}). One is established on the commutating operators that signifies the classical dependence between physical observables, the other is established on the nontrivial commutators from the BCH formula signifying the quantum dependence. Now we reformulate the GUR in a form that full orders of dependence appears coherently:
\begin{proposition}
For two physical observables, there exists a compatible uncertainty relation
\begin{align}
\langle e^{sX+s^*X}\rangle \langle e^{tY+t^*Y}\rangle \geq \left| \left\langle e^{sX+tY} \right\rangle \right|^2 \; , \label{Classical-UR}
\end{align}
and an incompatible one
\begin{align}
\langle e^{sX+s^*X}\rangle \langle e^{tY+t^*Y}\rangle \geq \left| \left\langle e^{sX+tY+\frac{1}{2}[sX,tY] + \cdots } \right\rangle \right|^2 \; . \label{Quantum-UR}
\end{align}
The former amounts to commutative classical observables, while the latter to noncommutative quantum observables. Here $e^{sX +tY + \frac{1}{2}[sX,tY]  +\cdots} =e^{sX}e^{tY} $, and the parameters $s$, $t$ can be arbitrary complex numbers. \label{Proposition-CQ}
\end{proposition}
The equations (\ref{Classical-UR}) and (\ref{Quantum-UR}) hold for all $s$ and $t$, and any observables. As equations (\ref{Classical-UR}) and (\ref{Quantum-UR}) are two special cases of the GUR, Proposition \ref{Proposition-CQ} indicates that the existence of uncertainty relation does not stem from the noncommutativity of the observables. This can also be seen from equation (\ref{Uncertainty-Sch}), where the bound on the right hand side may be nontrivial even if $[X,Y]=0$. This further rationalizes the ``dependence'' interpretation of the GUR.

\begin{figure}[t]\centering
\scalebox{0.5}{\includegraphics{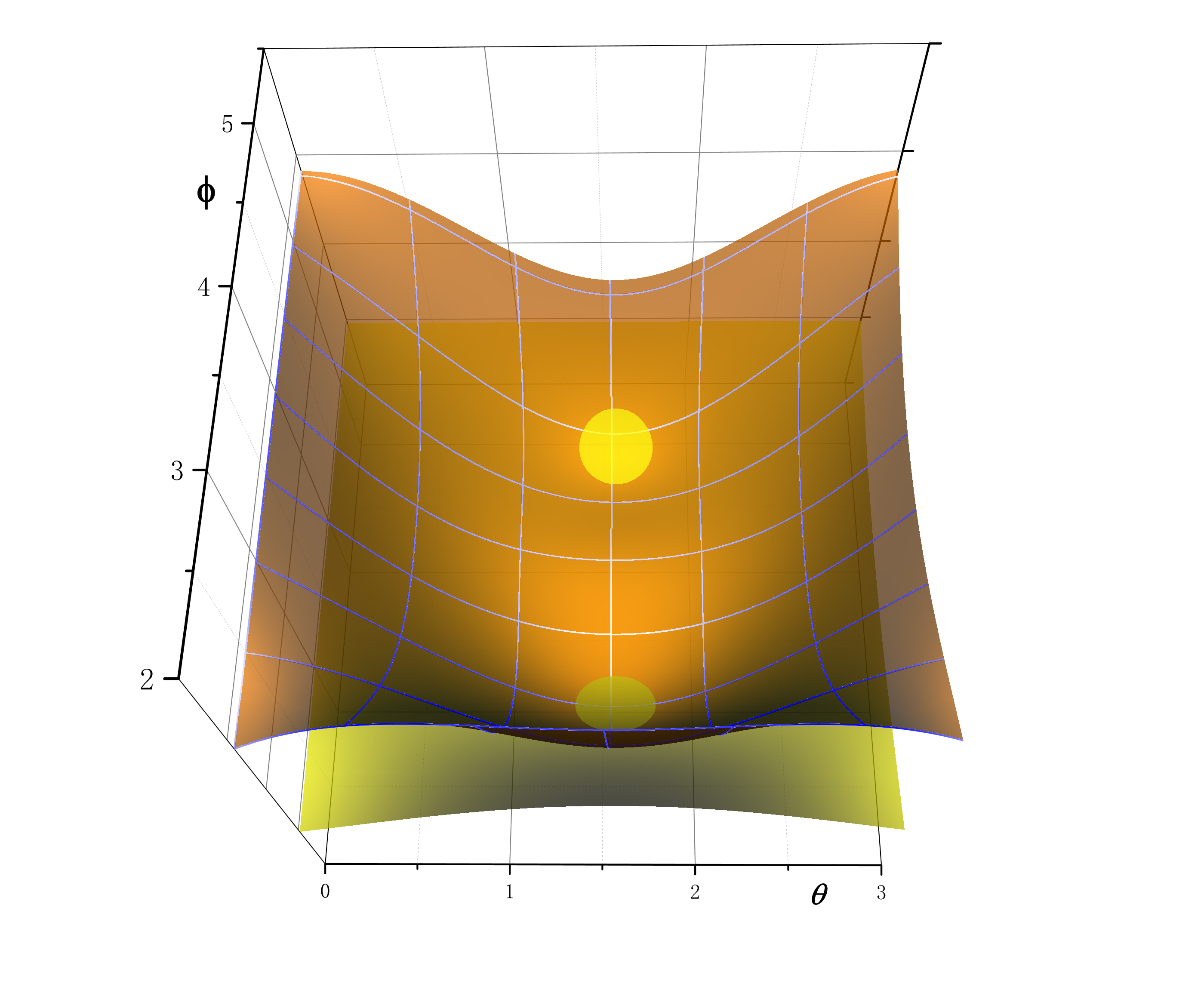}}
\caption{{\bf The quantum violation of the classical uncertainty relation.} The meshed surface represents the quantum prediction for the left hand side of relation (\ref{Classical-UR}), while the lower surface is the right hand side of it. The quantum state is taken to be $|\psi_1\rangle = \cos\frac{\theta}{2}|+\rangle + e^{i\phi}\sin\frac{\theta}{2} |-\rangle$ with two observables of $\sigma_x$ and $\sigma_y$. The violation happens in two circled regions around $(\theta,\phi) =\{(\frac{\pi}{2}, \pi), (\frac{\pi}{2}, \frac{3\pi}{2})\}$, where the two observables have statistical dependence that cannot be explained by classical theory.} \label{Figure-violation}
\end{figure}

\noindent{\bf Example 1} To exhibit the nonclassical dependence, we shall show the relation (\ref{Classical-UR}) may be violated in quantum theory through an example. For observables of $X=\sigma_x$ and $Y=\sigma_y$ with state $|\psi_1\rangle = \cos\frac{\theta}{2}|+\rangle + e^{i\phi}\sin\frac{\theta}{2} |-\rangle$, the left hand side of relation (\ref{Classical-UR}) is
\begin{align}
(\cosh2+\cos\phi\sin\theta\sinh2)(\cosh2+\sin\phi\sin\theta\sinh2) \;, \label{violation-left}
\end{align}
while the right hand side is
\begin{align}
\frac{1}{2}\left[ \sqrt{2}\cosh\sqrt{2}+ \sin\theta(\cos\phi+\sin\phi) \sinh\sqrt{2} \right]^2 \; . \label{violation-right}
\end{align}
Here, we have assumed $s=t=1$ for simplicity. Expressions (\ref{violation-left}) and (\ref{violation-right}) are numerically plotted in Figure \ref{Figure-violation}, where violations of equation (\ref{Classical-UR}) evidently exist. According to the generalized uncertainty principle, the two observables possess nonclassical statistical dependence in the violation region.

\noindent{\bf Example 2} There also exist the quantum constraints on two incompatible observables that can be detected by the uncertainty relation (\ref{Quantum-UR}), while failed by the uncertainty relation (\ref{Uncertainty-Sch}). We consider the two observables of angular momentums $L_x$ and $L_y$ where $[L_x,L_y]=i\hbar L_z \neq 0$. For the quantum state $|\psi_2\rangle = \frac{1}{\sqrt{3}} (|1\rangle + |0\rangle + |-1\rangle)$ of $L=1$ system, it is easy to find that the right hand of equation (\ref{Uncertainty-Sch}) is zero, which gives
\begin{align}
\Delta L_x^2 \Delta L_y^2  \geq 0 \; . \label{Example-sch}
\end{align}
Since the variance is greater than or equal to zero, (\ref{Example-sch}) tends to be trivial, which for example cannot rule out the possibility of $\Delta L_x^2 = \Delta L_y^2=0$, the precise measurements of $L_x$ and $L_y$ simultaneously. Nevertheless, for the same state and observables, the following equivalent form of equation (\ref{Quantum-UR}) for real $s$ and $t$
\begin{align}
\frac{\langle e^{sX+sX}\rangle}{|\langle e^{sX}\rangle|^2} \frac{\langle e^{tY+tY}\rangle}{|\langle e^{tY}\rangle|^2} & \geq \left| \frac{\left\langle e^{sX} e^{tY} \right \rangle}{\langle e^{sX}\rangle\langle e^{tY}\rangle} \right|^2 \; \label{Example-exp}
\end{align}
gives a nontrivial constraint. It is evident that for $s,t>0$ the right hand side of equation (\ref{Example-exp}) is always larger than 1, the accessible minimum of the left hand side. That means the (\ref{Example-exp}) may enforce certain constraint on the simultaneous measurements of $L_x$ and $L_y$. Therefore, we may conclude that in the framework of GUR $L_x$ and $L_y$ are linearly uncorrelated, but are higher order dependent. This is a novel insight regarding the compatibility of physical observables.

\noindent{\bf Example 3} The generalized uncertainty relation can be applied to the detection of entanglement as other uncertainty relations \cite{Ent-detect}. In qubit system, equation (\ref{W-uncertainty-2rd}) gives (see Appendix D)
\begin{align}
\kappa_2(\sigma_x) + \kappa_2(\sigma_y) + \kappa_2(\sigma_z) \geq 1 \; . \label{Example-xyz}
\end{align}
Here $\sigma_{i}$ are Pauli matrices. For local observables $A = \sigma_x\otimes \mathds{1} + \mathds{1} \otimes \sigma_x$,  $B = \sigma_y\otimes \mathds{1} + \mathds{1} \otimes \sigma_y$, and $C = \sigma_z\otimes \mathds{1} + \mathds{1} \otimes \sigma_z$ of two-qubit system, the separable states predicts  \cite{Ent-detect-G}
\begin{align}
\kappa_2(A) + \kappa_2(B) + \kappa_2(C) \geq 2 \; , \label{Example-ent}
\end{align}
where the lower bound $2$ is just the sum of the lower bound of equation (\ref{Example-xyz}) for both sides. While for the singlet state $|\psi_3\rangle = \frac{1}{\sqrt{2}}(|+-\rangle - |-+\rangle)$, we have
\begin{align}
\kappa_2(A) + \kappa_2(B) + \kappa_2(C) = 0\; .
\end{align}
This violates equation (\ref{Example-ent}) which indicates the non-separability of the state.

\noindent{\bf Example 4} The third order cumulant $\kappa_3$ can be used to exhibit new type of nonlocality \cite{Skew-nonlocal}. We consider the following operator for two-qubit system
\begin{align}
S = X \otimes Y - X \otimes Y' + X' \otimes Y + X'\otimes Y' \; .
\end{align}
Here we may choose $X= \sigma_z$, $X'=\sigma_x$, $Y = \sin\theta\sigma_x+\cos\theta\sigma_z$, and $Y' = \cos\theta\sigma_x-\sin\theta \sigma_z$. In classical correlated systems where $|\langle S \rangle| \leq 2$, there exists the following bound \cite{Skew-nonlocal}
\begin{align}
|\kappa_3(S) |= \left| \left\langle (S - \langle S \rangle )^3 \right\rangle \right| \leq 8 \; . \label{Example-nonlocal3}
\end{align}
The maximal value of $\kappa_3(S)$ in the singlet state $|\psi_3\rangle = \frac{1}{\sqrt{2}}(|+-\rangle - |-+\rangle)$ is $\frac{64\sqrt{6}}{9}$, which violates the equation (\ref{Example-nonlocal3}). This nonclassical phenomenon is called the ``skewness nonlocality'' \cite{Skew-nonlocal}. For the Werner state $\rho_{\mathrm{w}} = \frac{1-\eta}{4}\mathds{1} \otimes \mathds{1} + \eta|\psi_3\rangle \langle \psi_3|$ with $-\frac{1}{3} \leq \eta \leq 1$, it can be checked that the skewness nonlocality exists if $\eta>\displaystyle \frac{1}{\sqrt[3]{2}}-\frac{1}{3} \sim 0.46$. While the Bell nonlocality for projective measurements exists if and only if $\eta > \frac{1}{K_{G}(3)} \sim 0.66$, where $K_G(3)$ is the Grothendieck's constant of order three \cite{Bell-G-const}. Clearly, the skewness nonlocality is fundamentally different from the Bell nonlocality, and it arises purely from the higher order nonlinear dependence between the local observables in $\kappa_3(S)$.

\section{Conclusion}

We proposed a generalized uncertainty principle based on an alternative interpretation of Bohr's concept of complementarity. That is, the incompatibility between observables is interpreted as the statistical dependence between them. To elucidate the generalized principle, a GUR was derived which exhibits full-order statistical dependence between the physical observables. We found the dependence of observables can be well characterized by the statistical quantity of cumulant, by which the classical dependence turns out to be distinguishable from the quantum one. The second (first nontrivial) order approximation of GUR expansion yields the linear dependence and gives out the Heisenberg type uncertainty relation, while the third order skewness uncertainty relation predicts a constraint on the distribution asymmetries in different measurements. Concrete examples are also given to demonstrate the higher order dependence between observables and its possible applications in quantum information theory.

\section*{Acknowledgements}
\noindent
This work was supported in part by the National Natural Science Foundation of China under the Grants 11975236 and 11635009; and by the University of Chinese Academy of Sciences.

\newpage
\setcounter{figure}{0}
\renewcommand{\thefigure}{S\arabic{figure}}
\setcounter{equation}{0}
\renewcommand\theequation{S\arabic{equation}}
\setcounter{theorem}{0}
\renewcommand{\thetheorem}{S\arabic{theorem}}
\setcounter{observation}{0}
\renewcommand{\theobservation}{S\arabic{observation}}
\setcounter{proposition}{0}
\renewcommand{\theproposition}{S\arabic{proposition}}
\setcounter{lemma}{0}
\renewcommand{\thelemma}{S\arabic{lemma}}
\setcounter{corollary}{0}
\renewcommand{\thecorollary}{S\arabic{corollary}}

\appendix{\bf \Huge Appendix}

\section{Proof of Theorem 1}

For two physical obervables $X$ and $Y$, we consider two state vectors $e^{s^*X}|\psi\rangle$ and $e^{tY}|\psi\rangle$, where the Cauchy-Schwarz inequality tells
\begin{align}
\langle\psi |e^{sX} e^{s^*X} |\psi\rangle
\langle\psi |e^{t^*Y} e^{tY} |\psi\rangle \geq \langle \psi| e^{sX}e^{tY}|\psi\rangle \langle \psi| e^{sX}e^{tY}|\psi\rangle^* \; . \label{S-CS-inequality}
\end{align}
Both the left and right hand sides of the inequality (\ref{S-CS-inequality}) are positive definite. The logarithm function is monotone for positive real numbers, and we have
\begin{align}
\log\left[\langle e^{(s+s^*)X} \rangle \right] + \log\left[\langle e^{(t+t^*)Y} \rangle \right] \geq \log\left(\langle e^{Z_{st}} \rangle \right) + \log\left(\langle e^{Z_{st}} \rangle^* \right) \;.
\end{align}
Here $e^{Z_{st}} = e^{sX}e^{tY}$ and $Z_{st}= \log(e^{sX}e^{tY})=Z_{1} +Z_{11}+\cdots$. Using the cumulants generating functions $K[(s+s^*)X]= \log\left[\langle e^{(s+s^*)X} \rangle \right]$, $K[(t+t^*)Y]= \log\left[\langle e^{(t+t^*)Y} \rangle \right]$, and $K(Z_{st}) = \log\left(\langle e^{Z_{st}} \rangle \right)$, we arrive the Theorem 1.

Now we give a detailed expansion for terms in the generalized uncertainty relation (GUR)
\begin{align}
K[(s+s^*)X] + K[(t+t^*)Y] \geq K(Z_{st}) + \mathrm{c.c.} \; , \;  s, t \in \mathbb{C} \; , \label{S-K-uncertainty}
\end{align}
where ``c.c.'' means the complex conjugation of the previous term, i.e., $K^*(Z_{st})$. For $Z_{st}= \log(e^{sX}e^{tY})$, the BCH formula gives
\begin{eqnarray}
Z_{st}& =   & sX +tY + \frac{1}{2}[sX,tY] + \frac{1}{12}\left( [sX, [sX, tY]] + [tY, [tY, sX]] \right) \nonumber \\
& &  - \frac{1}{24}[tY, [sX, [sX,tY]]] - \nonumber \\
& & \frac{1}{720}\left( [[[[sX,tY],tY],tY],tY] + [[[[tY,sX],sX],sX],sX] \right) + \nonumber \\
& &  \frac{1}{360} \left( [[[[sX,tY],tY],tY],sX] + [[[[tY,sX],sX],sX],tY] \right) + \nonumber \\
& & \frac{1}{120}\left( [[[[tY,sX],tY],sX],tY] + [[[[sX,tY],sX],tY],sX] \right) +\cdots \nonumber \\
& = &  Z_{1}  + Z_{11} + Z_{21} + Z_{12} + Z_{22} + Z_{14}+Z_{41} + \nonumber \\
& &  Z^{(1)}_{23}+Z^{(1)}_{32} + Z^{(2)}_{23}+Z^{(2)}_{32} + \cdots \; . \label{S-Z-def}
\end{eqnarray}
The terms in the expansion of $Z_{st}$ represent
\begin{align}
Z_{1} & = sX + tY \; , \;
Z_{11} =  \frac{1}{2}[sX,tY] \; , \\
Z_{21} & = \frac{1}{12}[sX, [sX, tY]] \; , \;  Z_{12} =\frac{1}{12}  [tY, [tY, sX]] \; , \\
Z_{22} & = -\frac{1}{24}[tY, [sX, [sX,tY]]] \; , \\
Z_{14} & = -\frac{1}{720}  [[[[sX,tY],tY],tY],tY] \;, \;
Z_{41} = \frac{1}{720} [[[[tY,sX],sX],sX],sX]  \; , \\
Z_{23}^{(1)} & =  \frac{1}{360} [[[[sX,tY],tY],tY],sX] \; , \; Z_{32}^{(1)} = \frac{1}{360}  [[[[tY,sX],sX],sX],tY] \; , \\
Z_{23}^{(2)} &= \frac{1}{120} [[[[tY,sX],tY],sX],tY]  \; , \; Z_{32}^{(2)} = \frac{1}{120}  [[[[sX,tY],sX],tY],sX]  \; .
\end{align}
The left hand side of equation (\ref{S-K-uncertainty}) goes as
\begin{align}
K[(s+s^*)X] & = (s+s^*)\kappa_1(X) + \frac{(s+s^*)^2}{2!} \kappa_2(X) +  \frac{(s+s^*)^3}{3!} \kappa_3(X) +  \cdots \; , \label{S-expKx} \\
K[(t+t^*)Y] & = (t+t^*)\kappa_1(Y) + \frac{(t+t^*)^2}{2!} \kappa_2(Y) +  \frac{(t+t^*)^3}{3!} \kappa_3(Y) +  \cdots \; . \label{S-expKy}
\end{align}
While on the right hand side, we have
\begin{eqnarray}
K(Z_{st}) & = & \log\left(\left\langle \sum_{n=0}^{\infty}\frac{Z_{st}^n}{n!} \right\rangle \right) = \log\left(1+\langle Z_{st}\rangle + \frac{1}{2!}\langle Z_{st}^2\rangle + \frac{1}{3!}\langle Z_{st}^3\rangle+ \cdots \right) \nonumber \\
& = & \langle Z_{st} \rangle + \frac{1}{2!}\langle Z_{st}^2\rangle + \frac{1}{3!} \langle Z_{st}^3\rangle + \frac{1}{4!}\langle Z_{st}^4\rangle + \frac{1}{5!}\langle Z_{st}^5\rangle +\cdots  \nonumber \\
&  & -\frac{1}{2}\left(\langle Z_{st} \rangle + \frac{1}{2!}\langle Z_{st}^2\rangle + \frac{1}{3!} \langle Z_{st}^3\rangle + \frac{1}{4!}\langle Z_{st}^4\rangle +\cdots \right)^2 \nonumber \\
& & +\frac{1}{3} \left(\langle Z_{st} \rangle + \frac{1}{2!}\langle Z_{st}^2\rangle + \frac{1}{3!} \langle Z_{st}^3\rangle + \frac{1}{4!}\langle Z_{st}^4\rangle +\cdots \right)^3 \nonumber \\
& & -\frac{1}{4}\left(\langle Z_{st} \rangle + \frac{1}{2!}\langle Z_{st}^2\rangle + \frac{1}{3!} \langle Z_{st}^3\rangle + \frac{1}{4!}\langle Z_{st}^4\rangle +\cdots \right)^4 + \cdots \; .
\end{eqnarray}
Here $Z_{st}$ is defined in equation (\ref{S-Z-def}). For the second order, the coefficients of the terms $s^2$, $t^2$, and $st$ on right hand side can be read from the following
\begin{align}
K^{(2)}(Z_{st}) & = \langle Z_{11}\rangle + \frac{1}{2!}\langle Z_{1}^2 \rangle -\frac{1}{2} \langle Z_{1}\rangle^2 = \frac{\langle [sX,tY]\rangle}{2} + \frac{\langle (sX+tY)^2\rangle - \langle sX+tY\rangle^2}{2}\nonumber \\
& = \frac{st\langle [X,Y] \rangle}{2} + \frac{\kappa_{2}(sX)+  \kappa_{2}(tY)+ 2\kappa_{11}(sX,tY)}{2} \;. \label{S-expKz}
\end{align}
For the third order, the coefficients of $s^3$, $t^3$, $s^2t$, and $st^2$ on the right hand side are
\begin{eqnarray}
K^{(3)}(Z_{st}) & = & \langle Z_{12}+Z_{21} \rangle + \frac{\langle Z_{1}Z_{11}+Z_{11}Z_{1}\rangle }{2!} + \frac{ \langle Z_{1}^3 \rangle}{3!}  -\frac{\langle Z_{1} \rangle \displaystyle \left(\langle Z_{1}^2\rangle +2 \langle Z_{11}\rangle\right)}{2}  +\frac{\langle Z_{1}\rangle^3}{3!} \nonumber \\
& = & \frac{\langle[sX,[sX,tY]]\rangle+\langle [tY,[tY,sX]] \rangle}{12} + \frac{ \langle \{sX+tY, [sX,tY]\}\rangle}{4} + \frac{\langle (sX+tY)^3\rangle}{3!}\nonumber \\
& & -\frac{\langle sX+tY\rangle(\langle (sX+tY)^2\rangle + \langle [sX,tY]\rangle)}{2} + \frac{\langle sX+tY\rangle^3}{3} \; ,
\end{eqnarray}
which may be simplified into
\begin{align}
K^{(3)}(Z_{st}) =& \frac{\langle[sX,[sX,tY]]\rangle+\langle [tY,[tY,sX]] \rangle}{12} + \frac{ \langle \{sX+tY, [sX,tY]\}\rangle}{4} \nonumber \\
&  -\frac{\langle sX+tY\rangle \langle [sX,tY]\rangle}{2} + \frac{\kappa_{3}(sX+tY)}{3!} \nonumber \\
 = &  \frac{\langle[sX,[sX,tY]]\rangle+\langle [tY,[tY,sX]] \rangle}{12} + \frac{ \langle \{sX+tY, [sX,tY]\}\rangle -2\langle sX+tY\rangle\langle [sX,tY]\rangle}{4} \nonumber \\
 &  + \frac{\kappa_{3}(sX+tY)}{3!}\;.
\end{align}
Here $\kappa_3(sX+tY) = \kappa_{3}(sX)+ 3\kappa_{12}(sX,tY)+3\kappa_{21}(sX,tY)+\kappa_{3}(tY)$.

\section{Proof of Corollary 2}

The equation (20) in Corollary 2 is quite straightforward from the expansions of equations (\ref{S-expKx}), (\ref{S-expKy}), and (\ref{S-expKz}). Because for $|s| \sim |t| \ll 1, s,t \in \mathbb{C}$, we have
\begin{align}
|s|^2 \kappa_2(X) + |t|^2 \kappa_2(Y) \geq & \left( st\frac{\langle XY+YX\rangle -2\langle X\rangle\langle Y\rangle}{2} + st\frac{\langle [X,Y]\rangle}{2} \right) + \mathrm{c.c.} \nonumber \\
= & \mathrm{Re}[st] (\langle XY+YX\rangle -2\langle X\rangle\langle Y\rangle) + \mathrm{Im}[st]i\langle [X,Y]\rangle \; , \label{S-k2-uncert}
\end{align}
where ``c.c.'' means the complex conjugate of the previous term in the bracket; ``Re'' and ``Im'' stand for the real and image parts of the parameters. The Cauchy-Schwarz inequality further implies
\begin{align}
& \mathrm{Re}[st] (\langle XY+YX\rangle -2\langle X\rangle\langle Y\rangle) + \mathrm{Im}[st]i\langle [X,Y]\rangle \nonumber \\
& \leq \sqrt{\mathrm{Re}(st)^2+\mathrm{Im}(st)^2} \sqrt{|\langle XY+YX\rangle -2\langle X\rangle\langle Y\rangle|^2 + |\langle [X,Y]\rangle|^2} \nonumber \\
& = |st| \sqrt{|\langle XY+YX\rangle -2\langle X\rangle\langle Y\rangle|^2 + |\langle [X,Y]\rangle|^2} \;, \label{S-k2-final}
\end{align}
and the equality is satisfied for appropriately chose real and image parts of $st$. In the case of the equality being satisfied, the right hand side of equation (\ref{S-k2-final}) becomes the right hand side of equation (\ref{S-k2-uncert}), and equation (21) in the main text is arrived.

\section{The truncated cumulant expansions}

The power series of $K(sX)$ can be written as
\begin{align}
K(sX) = &  \log (\langle e^{sX}\rangle) = \log\left(1 + \langle sX\rangle + \frac{s^2}{2!}\langle X^2\rangle + \frac{s^3}{3!}\langle X^3\rangle+ \cdots \right) \nonumber \\
= & f(s) -\frac{f^2(s)}{2} + \frac{f^3(s)}{3} - \frac{f^4(s)}{4} +\cdots \; .
\end{align}
Here $f(s) = \langle sX\rangle + \frac{s^2}{2!}\langle X^2\rangle + \frac{s^3}{3!}\langle X^3\rangle+ \cdots = \langle e^{sX}\rangle -1$ with $f(0)=0$. The convergence condition for $f(s)$ around $f(s)=0$ is
\begin{align}
\left| f(s) -0 \right| = \left|\langle \psi| e^{sX}| \psi\rangle -1 \right| <1 \;,\; \forall\; |\psi\rangle . \label{S-conv-condition1}
\end{align}
Consider the special case of $s\in \mathbb{R}$, we have the convergence region for $s$ around $s=0$
\begin{align}
0 < \langle e^{sX} \rangle <2 \Longrightarrow -\frac{\log 2}{\sigma_{\mathrm{max}}} <s  <\frac{\log 2}{\sigma_{\mathrm{max}}}\; ,
\end{align}
where $\sigma_{\mathrm{max}}$ is the largest singular value of Hermitian matrix $X$. For continuous observable $X$ whose probability distribution has the typical width of $\sigma$ around $x_0$, we have
\begin{align}
e^{\frac{s^2\sigma^2}{2}+sx_0} <2 \Longrightarrow \frac{-\sqrt{2\sigma^2\log2 + x_0^2}-x_0}{\sigma^2}< s < \frac{\sqrt{2\sigma^2\log2 + x_0^2}-x_0}{\sigma^2} \; .
\end{align}
Here we have assumed that the moments $\langle X^n\rangle$ exist for all orders $n$. Approximately, there exist
\begin{eqnarray}
x_0 \ll \sigma & : &
\frac{-\sqrt{2\log2}}{\sigma}< s < \frac{\sqrt{2\log2}}{\sigma} \;,  \\
\sigma \ll x_0 & : &
\frac{-\log2}{|x_0|}< s < \frac{\log2}{|x_0|} \;.
\end{eqnarray}

Now we study some of the trivial implications of equation (26) for infinitesimal real values of $s\sim t \to \varepsilon$. For independent observables, the right hand side of equation (26) is zero, and by expressing the cumulants in terms of moments we have
\begin{align}
& \kappa_2(X) + \kappa_2(Y) +  \varepsilon \left[ \kappa_3(X)+ \kappa_3(Y) \right] \nonumber \\
= & \langle (X-\langle X\rangle)^2\rangle + \langle (Y-\langle Y\rangle)^2\rangle + \varepsilon \left[ \langle (X-\langle X\rangle)^3\rangle + \langle (Y-\langle Y\rangle)^3\rangle \right] \geq 0 \; . \label{S-trivial}
\end{align}
Or equivalently,
\begin{align}
\left\langle (X-\langle X\rangle)^2 \left[1 + \varepsilon (X-\langle X\rangle)\right] + (Y-\langle Y\rangle)^2 \left[1+ \varepsilon (Y-\langle Y\rangle) \right] \right\rangle \geq 0 \;.
\end{align}
To ensure the convergence of both $K(2\varepsilon X)$ and $K(2\varepsilon Y)$, the real parameter $\varepsilon$ shall be
\begin{align}
2|\varepsilon |\sigma_{\mathrm{max}}< \log 2 \; .
\end{align}
Here $\sigma_{\mathrm{max}}$ is the largest singular value for $X$ and $Y$. The spectrums of $(X-\langle X\rangle)$ and $(Y-\langle Y\rangle)$ lie in $[-2\sigma_{\mathrm{max}}, 2\sigma_{\mathrm{max}}]$, so the minimal eigenvalue $\lambda_{\mathrm{min}}$ of $1+\varepsilon (X-\langle X\rangle)$ and $1+\varepsilon (Y-\langle Y\rangle)$ is
\begin{align}
\lambda_{\mathrm{min}} > 1-2|\varepsilon|\sigma_{\mathrm{max}} >1-\log2 >0 \; .
\end{align}
The equation (\ref{S-trivial}) is trivially right for independent variables, i.e.,
\begin{align}
\kappa_2(X) + \kappa_2(Y) + \varepsilon \left[ \kappa_3(X)+ \kappa_3(Y) \right] \geq 0 \;.
\end{align}

\section{Derivations of the Examples}

\noindent{\bf Example 3} In qubit system, equation (20) predicts that the sum of the second order cumulants of the observables pairs $(\sigma_x,\sigma_y)$, $(\sigma_y,\sigma_z)$, and $(\sigma_z,\sigma_x)$ are
\begin{align}
& \hspace{0.2cm} |s_1|^2\kappa_{2}(\sigma_x) + |t_1|^2\kappa_{2}(\sigma_y) + |s_2|^2\kappa_{2}(\sigma_y) + |t_2|^2\kappa_{2}(\sigma_z) + |s_3|^2\kappa_{2}(\sigma_z) + |t_3|^2\kappa_{2}(\sigma_x) \nonumber \\
\geq & \hspace{0.2cm} 2\left( |s_1t_1|(|\langle \sigma_x\rangle\langle \sigma_y\rangle|^2+ |\langle \sigma_z\rangle|^2)^{\frac{1}{2}} + |s_2t_2|(|\langle \sigma_y\rangle\langle \sigma_z\rangle|^2+ |\langle \sigma_x\rangle|^2)^{\frac{1}{2}}  \right. \nonumber \\ & \left. + |s_3t_3| (|\langle \sigma_z\rangle\langle \sigma_x\rangle|^2 + |\langle \sigma_y\rangle|^2)^{\frac{1}{2}} \right) \; , \label{S-Example-ent-det}
\end{align}
where we have used the anti-commutativity of the Pauli operators and appropriately choosed phases of $s_i$ and $t_i$ as that in Appendix B. If we set $|s_1|=|t_1|=|s_2|=|t_2|=|s_3| = |t_3|$, then
\begin{align}
\kappa_2(\sigma_x) + \kappa_2(\sigma_y) + \kappa_2(\sigma_z) \geq \zeta_1+\zeta_2 + \zeta_3 \geq 1 \; .
\end{align}
Here $\zeta_1 = (|\langle \sigma_x\rangle \langle \sigma_y\rangle|^2 + |\langle \sigma_z\rangle|^2)^{\frac{1}{2}}$, $\zeta_2 = (|\langle \sigma_y\rangle \langle \sigma_z\rangle|^2 + |\langle \sigma_x\rangle|^2)^{\frac{1}{2}}$, and $\zeta_3 = (|\langle \sigma_z\rangle \langle \sigma_x\rangle|^2 + |\langle \sigma_y\rangle|^2)^{\frac{1}{2}}$

Generally, we may set $|s_1t_1|=\varepsilon_1^2$, $|s_2t_2|=\varepsilon_2^2$, $|s_3t_3|=\varepsilon_3^2$ where $\varepsilon_{i}$ are real parameters. The right hand side of equation (\ref{S-Example-ent-det}) can be made as large as
\begin{align}
2\sqrt{(\varepsilon_{1}^4+\varepsilon_{2}^4+\varepsilon_{3}^4) (\zeta_{1}^2 + \zeta_2^2 + \zeta_3^2 )} \; ,
\end{align}
at the condition of $\displaystyle \frac{\varepsilon_i^2}{\sqrt{\varepsilon_{1}^4 +\varepsilon_{2}^4 + \varepsilon_{3}^4}} = \frac{\zeta_i}{\sqrt{\zeta_{1}^2 + \zeta_2^2 + \zeta_3^2 }}$, $i=1,2,3$.
Equation (\ref{S-Example-ent-det}) now can be reexpressed as
\begin{align}
\zeta_1 \left[ \frac{|s_1|}{|t_1|}|\kappa_2(\sigma_x) + \frac{|t_1|}{|s_1|}|\kappa_2(\sigma_y) \right]+
\zeta_2 \left[ \frac{|s_2|}{|t_2|}|\kappa_2(\sigma_y) + \frac{|t_2|}{|s_2|}|\kappa_2(\sigma_z) \right] \nonumber \\ + \zeta_3 \left[ \frac{|s_3|}{|t_3|}|\kappa_2(\sigma_z) + \frac{|t_3|}{|s_3|}| \kappa_2(\sigma_x) \right]  \geq 2 \sqrt{\zeta_1^2+ \zeta_2^2 + \zeta_3^2} \; .
\end{align}
This gives a variance uncertainty relation for qubit states.

\noindent{\bf Example 4} The quantum mechanical prediction for $S = X\otimes Y - X\otimes Y' + X'\otimes Y + X'\otimes Y'$ is
\begin{align}
\langle S\rangle = E(X,Y)-E(X,Y') + E(X',Y) + E(X',Y')\; ,
\end{align}
where the correlation is evaluated as $E(X,Y) = \langle X\otimes Y\rangle$, and $X= \sigma_z$, $X'=\sigma_x$, $Y = \sin\theta\sigma_x+ \cos\theta\sigma_z$, and $Y' = \cos\theta\sigma_x-\sin\theta \sigma_z$. In the local hidden variable theories (LHVTs), there is the following inequality related to $S$
\begin{align}
S= A(\lambda,X)[B(\lambda,Y)-B(\lambda,Y')] + A(\lambda,X')[B(\lambda,Y)+B(\lambda,Y')] \in [-2,2]\; .
\end{align}
Here $\langle S\rangle \equiv \displaystyle \int \xi_{\lambda} S\,\mathrm{d}\lambda$ and the correlation becomes $E(X,Y) =  \displaystyle\int \xi_{\lambda} A(\lambda,X)B(\lambda,Y) \, \mathrm{d}\lambda$ with $\xi_{\lambda}$ being an unknown distributions of hidden variable(s). For random variable $-2\leq S \leq 2$, there is
\begin{align}
\left| \kappa_3(S) \right| & = \left| \int \left(S-\langle S\rangle \right)^3 \xi_{\lambda} \mathrm{d}\lambda \right| \leq \int |S-\langle S\rangle|^3 \xi_{\lambda} \,\mathrm{d} \lambda \leq 8 \; , \label{S-k38}
\end{align}
where we have used the fact that, for $-2 \leq S \leq 2$, the 3th central moment is less than $8$ \cite{Cent-Moment}.

The two-qubit Werner state is defined as
\begin{align}
\rho_{\mathrm{w}} \equiv \frac{1-\eta}{4}\mathds{1} \otimes \mathds{1} + \eta|\psi_3\rangle \langle \psi_3| \; .
\end{align}
Here $|\psi_3\rangle = \frac{1}{\sqrt{2}}(|+-\rangle - |-+\rangle)$ and $\eta \in [-1/3,1]$. The quantum mechanical result of $\kappa_3(S)$ for $\rho_{\mathrm{w}}$ is
\begin{align}
\kappa_3(S) = -8\eta (\cos\theta+\sin\theta)[-1-3\eta+ 2\eta^2(\cos\theta+\sin\theta)^2]\; .
\end{align}
It can be checked that for singlet state of $\eta=1$, $\kappa_3(S)$ may reach a maximal value of $\frac{64\sqrt{6}}{9}\sim 17.42>8$. The violation of equation (\ref{S-k38}) remains for $\eta > \displaystyle \frac{1}{\sqrt[3]{2}}-\frac{1}{3} \sim 0.46$.

\end{document}